\documentclass[twocolumn,showpacs,preprintnumbers,amsmath,amssymb]{revtex4}
\usepackage[dvips]{graphicx}
\usepackage{dcolumn}
\usepackage{bm}

\begin{document}

\title{Thermal budget of superconducting digital circuits at
sub-kelvin temperatures}

\author{A.M. Savin}
\author{J.P. Pekola}
\affiliation{Low Temperature Laboratory, Helsinki University of
Technology, P.O. Box 3500, 02015 HUT, Finland}

\author{D.V. Averin}
\author{V.K. Semenov}
\affiliation{Department of Physics and Astronomy, Stony Brook
University, Stony Brook, New York 11974-3800, USA}


\begin{abstract}
Superconducting single-flux-quantum (SFQ) circuits have so far
been developed and optimized for operation at or above helium
temperatures. The SFQ approach, however, should also provide
potentially viable and scalable control and read-out circuits for
Josephson-junction qubits and other applications with much lower,
milli-kelvin, operating temperatures. This paper analyzes the
overheating problem which becomes important in this new
temperature range. We suggest a thermal model of the SFQ circuits
at sub-kelvin temperatures and present experimental results on
overheating of electrons and silicon substrate which support this
model. The model establishes quantitative limitations on the
dissipated power both for ``local'' electron overheating in
resistors and ``global'' overheating due to ballistic phonon
propagation along the substrate. Possible changes in the thermal
design of SFQ circuits in view of the overheating problem are also
discussed.

\end{abstract}

\maketitle

\section{Introduction}

It is widely accepted that potential scalability by means of the
present-day or prospective microelectronic technology is the main
advantage of solid state qubits, and in particular,
superconducting qubits (see, e.g., \cite{q1,q2}). One of the many
requirements necessary to realize this potential is a reasonably
high integration density of both the qubit and control circuits,
which almost unavoidably means that control circuits should be
located close to qubits with their milli-kelvin operating
temperatures, and are allowed to dissipate only small amount of
energy. The requirement of low energy dissipation and ability to
function below liquid-helium temperatures make superconductor
Single-Flux-Quantum (SFQ) devices \cite{rsfq} the most promising
candidate for prospective qubit control circuit technology
\cite{sfq1}. Reported SFQ devices are also much faster than their
semiconductor counterparts (see, e.g., \cite {sfq2}) and, as a
result, should provide a much better accuracy of qubit control.

Although the SFQ circuits have been investigated for many years,
one of the implied ``design objectives'' of these investigations
was the possibility to increase rather than decrease operating
temperature, and many of the suggested approaches can not be
immediately applied to qubit control circuits. The main new
obstacle introduced by low operating temperatures is a dramatic
degradation of thermal conductivities of all materials at
milli-kelvin temperatures. This should cause strong overheating of
the SFQ circuits, which dissipate power that is small in
comparison to semiconductor devices, but is still very significant
in the sub-kelvin temperature range. Overheating establishes
effective temperature of the SFQ components far above the bath
temperature. It also affects the qubit part of the circuit both
directly, through the heat flow to qubits, and indirectly, by
creating stronger electromagnetic noise that acts as an extra
source of decoherence for qibits.

In this work, we analyze the overheating problem facing SFQ
circuits at sub-kelvin temperatures. The analysis results in
semi-quantitative understanding of the magnitude of the
SFQ-induced disturbance of the qubits, and re-scaling of the SFQ
circuits required to satisfy thermal constraints of the sub-kelvin
temperature range. The main elements of this re-scaling can be
summarized as follows. For a given clock frequency of a
conventional SFQ circuit, its power dissipation $P$ is
proportional to typical critical current of the Josephson
junctions in the circuit. In its turn, the critical current can
not be reduced below some thermal value which scales linearly with
effective operating temperature $T$ of the circuit because of the
thermally induced errors in its dynamics. Finally, the temperature
$T$ is determined by the balance between the dissipated power $P$
and efficiency of the heat removal from the circuit.
Qualitatively, since the thermal conductivities of all materials
show strong dependence on the temperature $T$, direct re-scaling
of conventional SFQ circuits will be capable of providing only
relatively modest reduction of their effective temperature (in
practical terms, to about 0.4 K). Overheating of the SFQ
components of this magnitude requires their careful thermal
insulation from the qubits, which in the case of SFQ circuits with
large complexity can be easily achieved by placing them on a
separate chip. Alternative solutions, such as specially modified
substrates or advanced thermal coupling with the sink, are more
complicated and as a result they could be recommended for
"industrial type" projects.

\section{Heat Flow at Sub-Kelvin Temperatures and Estimates of the Thermal
Resistances}

\begin{figure}[t]
\includegraphics[width=0.9\linewidth]{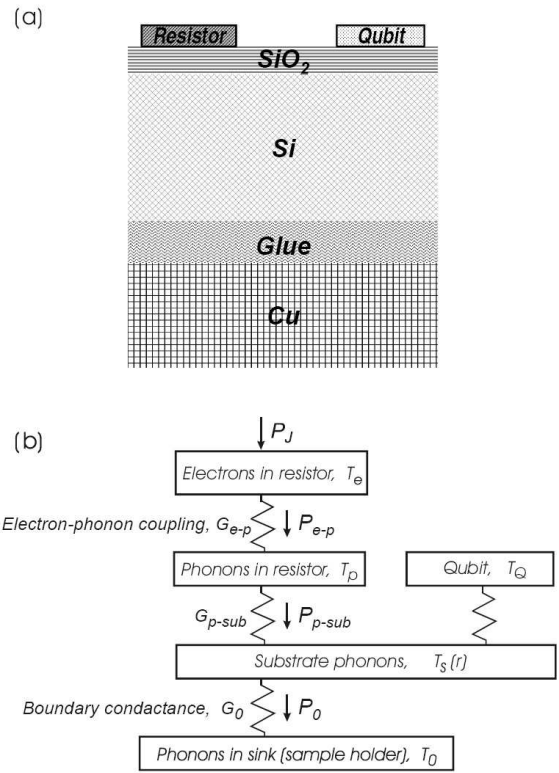}
\caption{Cross section (a) and a simplified thermal diagram (b) of
a typical SFQ-qubit integrated circuit glued to a massive thermal
sink.} \label{CrossCecab}
\end{figure}

Temperature of a superconductor integrated circuit is defined by
the balance of the power dissipated in the circuit resistances and
the efficiency of transfer of this power from the circuit to the
sink. We start by discussing the thermal conductance between
heat-generating resistors and the heat sink. A typical thermal
structure of a superconductor circuit is shown in
Fig.~\ref{CrossCecab}. Its complexity makes precise determination
of temperature distribution in prospective SFQ and qubit circuits
hardly possible. There are several factors that have the main
effect on the temperature distribution. The first is the
electron-phonon coupling limiting the heat transfer between
electron gas and the lattice which is responsible for the electron
overheating. The magnitude of the lattice overheating is mainly
determined by the competition of the thermal resistance associated
with the phonon propagation along the substrate on the one hand,
and on the other hand, the boundary (``Kapitza'') resistances
between the adjacent layers of different materials due to their
acoustic mismatch or the thermal resistance of amorphous
dielectrics (typically SiO$_2$ and different epoxies) used for
electric isolation of circuit components and for thermal
connections of the integrated circuit with the sink.

Phenomenologically, all heat transfer mechanisms should be rather
similar at low temperatures, and are characterized by the
material-independent power $\beta$ of the power-law dependence of
the heat flux on temperature, a material-dependent prefactor
$\gamma$ in this power-law \cite {lounasmaa74, berman, pobell92}.
In the "differential" form, expression for the heat flux $P$
between the two regions with temperatures $T_{1}$ and $T_{2}$ is
valid for small temperature difference $\Delta T\equiv
T_{2}-T_{1}<<T_{1} \simeq T_{2}\equiv T$:
\begin{equation}
P=\gamma T^{\beta}\Delta T \, .  \label{PowDep}
\end{equation}%
The corresponding "integral" expression valid for arbitrary $T_{1}$ and $%
T_{2}$ is
\begin{equation}
P=\frac{\gamma }{\beta +1}(T_{2}^{\beta +1}-T_{1}^{\beta +1}) \, .
\label{PowDep2}
\end{equation}

In the next subsections we discuss different specific mechanisms
of the heat conduction and their effect on the temperature
distribution in the SFQ circuits.

\subsection{Electron-phonon coupling}

Electrons in the bias and shunt resistors are the main sources of
the dissipated energy in the SFQ circuits, and as a result, have
the highest temperature among the elements of the circuit. This
temperature is the most important one for the circuit operation
since the magnitude of the fluctuation-induced errors obviously
depends on the electron rather than phonon temperature. Resistors
in the SFQ circuits are typically attached at the ends to
superconducting electrodes so that the heat flow through the
contacts is suppressed by Andreev refection. Electron-phonon
relaxation provides then the main mechanism of electron cooling in
the resistors. According to the standard model of this relaxation
in a metal, steady-state electron temperature $T_{e}$ and the
lattice temperature $T_p$ in the resistor are related as follows
\cite{roukes85, wellstood:5942}:
\begin{equation}
P_{e-p}=\Sigma \Lambda (T_{e}^{5}-T_{p}^{5}). \label{ElTemp}
\end{equation}
Here $P_{e-p}$ is the heat flux between the electrons and the
lattice, $\Sigma$ is a material constant, and $\Lambda$ is the
volume of the resistor. For metals, typical value of $\Sigma$ is
$\Sigma \simeq 10^{9}$ Wm$^{-3}$K$^{-5}$. Equation (\ref{ElTemp})
shows that electron-phonon coupling decreases very rapidly with
temperature, and at sub-kelvin temperatures electrons in the
resistors are significantly overheated by electrical current.
Because of the strong power-law dependence in Eq.~(\ref{ElTemp}),
and for power values relevant for the SFQ circuits, electron
temperature is determined mostly by the applied power and only
little by the lattice temperature and for $T_{e}>T_{p}$ can be
safely estimated as $T_{e}\simeq (P_{J}/(\Sigma \Lambda ))^{1/5}$,
where $P_{J}$ is the Joule heating due to electrical current
through the resistor. For the power range and resistor volumes  of
interest (on the order of nW and $\mu$m$^3$, respectively), this
estimate falls into the temperature interval 0.1 K - 1 K. In this
temperature range, the thermal resistance ($G^{-1}_{p-sub}$)
between phonons in the film and the substrate is usually taken to
be rather small compared to the electron-phonon resistance
($G^{-1}_{e-p}$) due to strong coupling between phonon systems in
a thin film and a substrate \cite{roukes85, wellstood:2599,
echternach:10339}, so that the effective ``electron-substrate''
coupling can be described by Eq.~(\ref{ElTemp}).

\begin{figure}[t]
\includegraphics[width=0.8\linewidth]{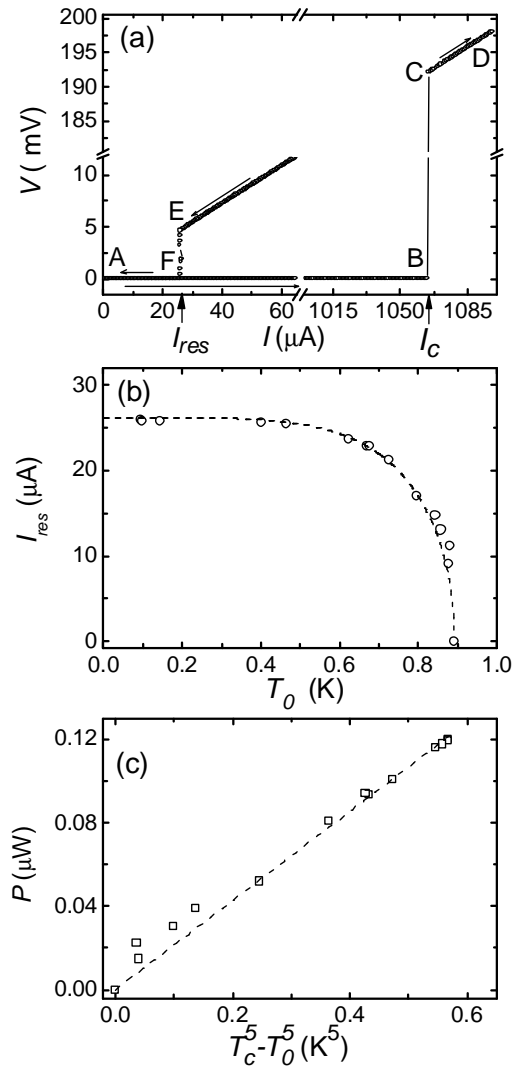}
\caption{Electron-phonon coupling experiment (see details in the
text). (a) Transition of a Mo resistor from superconducting to
normal state (A $\rightarrow$ B $\rightarrow$ C $\rightarrow$ D)
and from normal to superconducting state (D $\rightarrow$ E
$\rightarrow$ F $\rightarrow$ A) at $T_{0}$ = 380 mK. (b)
Dependence of $I_{\mathrm{res}}$ on bath temperature $T_{0}$. (c)
Dependence of heating power on $T_{\mathrm{e}}^{5}-T_{0}^{5}$.}
\label{MoRes1}
\end{figure}

An experiment to check the validity of Eq.~(\ref{ElTemp}) and the
arguments above in realistic conditions was performed, using a
typical wafer from HYPRES \cite{HYPRES}, Inc., with molybdenum
resistors. The idea of the experiment was to use transition at
temperature $T_{c}$ of Mo resistor from normal to superconducting
state as an electron thermometer and estimate the electron-phonon
thermal coupling as follows. In spite of the limitation that we
can detect only one electron temperature $T_{c}$ (measured to be
$\simeq 0.893$ K for these Mo resistors) we can measure the Joule
heating power needed to keep the resistor in the resistive state
at a temperature just above $T_{c}$, for different (smaller)
lattice temperatures. Thus this measurement produces all the data
necessary for analysis based on Eq.~(\ref{ElTemp}): the electron
temperature $T_{e}=T_{c}$, the heating power, and the phonon
temperature which we take to be equal to the temperature $T_{0}$
of the sample holder. This approximation is justified, since at
temperatures $T_{0}$ below the electron temperature $T_{c}$ the
actual phonon temperature is practically irrelevant because of the
strong power-law dependence in Eq.~(\ref{ElTemp}).

Figure \ref{MoRes1}(a) illustrates our measurement procedure. We
make use of strong hysteresis of the current-voltage
characteristics of the resistor. Initially, at low currents, the
resistor remains in the superconducting state and $P_{J}\equiv 0$
(trace A $\rightarrow $ B in Fig. \ref{MoRes1}(a)). Only after
exceeding the critical current of the molybdenum strip ($I_{c}$ =
1.065 mA at $T_{0}$ = 380 mK in Fig. \ref{MoRes1}(a)) we dissipate
power in the resistor (B $ \rightarrow $ C corresponds to
superconducting - normal state transition). Now the Joule heating
is very strong, and only by reducing the current far below the
critical current to $I_{res}(T_{0})$ (D $\rightarrow $ E in Fig.
\ref{MoRes1}(a)) we reduce the electron temperature sufficiently
to finally detect transition back to the superconducting state (E
$\rightarrow $ F). The power at this working point,
$P_{J}=RI_{res}^{2}(T_{0})$ heats the system up to $T_{e}=T_{c}$.
Experimental values of $I_{res}$ at different bath temperatures
and heating power $P_{J}$ as a function of $T_{e}^{5}-T_{0}^{5}$
are presented in Fig. \ref{MoRes1}(b,c). The dashed line shows
good fit of the measured data using Eq.~(\ref{ElTemp}), $\Lambda
=$ 24 $\mu $m$^{3}$, and
\begin{equation}
\Sigma =0.9\cdot 10^{9}\text{ Wm$^{-3}$K$^{-5}$.} \label{Sigma}
\end{equation}
We see that the electron-phonon constant obtained from this fit is
indeed in line with the typical metal values.

Generally an SFQ circuit contains a large number of resistors with
different electronic temperatures and the error rate depends on
all these temperatures. However, the cumulative effect of
electronic temperatures $T_{eb}$ and $T_{es}$ of the two resistors
used to bias ($R_{b}$) and shunt ($R_{s}$) the same Josephson
junction can be reduced to a single noise temperature $T_{N}$:
\begin{equation}
T_{N}=(T_{eb} R_{s}+T_{es} R_{b})/(R_{s}+R_{b}).
\label{EffNoiseTemp}
\end{equation}
To get a feeling for the magnitude of possible electron
overheating we estimate the noise temperature $T_{N}$ for a
typical junction with critical current $I_{c}= $ 10 $ \mu$A biased
by the dc current $I_{b}$ = 7 $\mu$A. The junction is critically
damped ($\beta _{c}= 1$) by a resistor $R_{s} = $ 10 $\Omega$
($I_{c}R_{s}$ = 100 $\mu$V) and the bias voltage $V_{b}$ is taken
to be about 300 $\mu $V, i.e., $ R_{b}=$ 43 $\Omega$. These
parameters are reasonable for the fabrication technology for the
sub-kelvin circuits offered by HYPRES, which is the only one
available commercially. In this technology, Josephson junctions
have 100 A/cm$^{2}$ density of critical current and all resistors
are made of 0.1 $\mu $m $PdAu$ film with 2 $\Omega$ sheet
resistance. The bias current $I_{b}$ in the resistor $R_{b}$ is
nearly time-independent and therefore $T_{eb}$ can be estimated in
the steady-state model. Resistor volume $\Lambda _{b}$ required
for the calculation of the specific heat flux $P_{e-p}$ can be
varied freely while keeping its resistance (set by the ratio of
its length and width) constant. According to a conventional
(miniaturization) wisdom the resistor dimensions should be made as
small as possible. In HYPRES technology, the minimal recommended
dimension (width) is 3 $\mu $m, and 43 $\Omega$ resistor is 65
$\mu $m long in this case. For comparison we will calculate also
overheating of the bias resistor of a larger size, with 10 $\mu $m
width.

From the dissipated power $P_{J}$ = 7 $\mu $A $\cdot$ 300 $\mu
$V$=2.1\cdot 10^{-9}$ W and resistor volume $\Lambda _{b}$ that
for our two examples is equal to $ 1.9\cdot 10^{-17}$ m$^{3}$ and
$2.1\cdot 10^{-16}$ m$^{3}$, we see that at $T_{p}=0$ the
corresponding electronic temperatures in the two cases are very
close: 0.64 K and 0.4 K, despite the factor-of-10 difference in
volumes. Shunt resistor $R_{s}$ can in principle be much colder
(with $T_{e}$ approaching $T_{p}$) since the current flows via it
only during short (pico-second) SFQ pulses generated by the
Josephson junction. Each of such pulses dissipates energy of about
$I_{c}\Phi _{0}$, where $\Phi _{0}$ is the magnetic flux quantum,
$\Phi _{0}=\pi \hbar /e$. Corresponding evolution of electron
temperature is described by the equation:
\begin{equation}
P_{J}=P_{e-p}+C_{e} dT_{e}/dt,  \label{HeatBal}
\end{equation}
where the heat flux $P_{e-p}$ is given by Eq.~(\ref{ElTemp}), and
$C_{e}$ is the heat capacity of the electron gas. The linear
dependence of $C_{e}$ on temperature: $C_{e}=\gamma _{e}T_{e}$,
where $\gamma _{e}\simeq 200$ J m$^{-3}$ K$^{-2}$ \cite{kittel},
and the fact that in the relevant temperature range the SFQ pulses
are very fast in comparison to the relaxation time
\begin{equation}
\tau _{e-p} = (\gamma _{e}/3\Sigma)\cdot T^{-3}  \simeq (0.07
\text{ $\mu$s K$^{3}$}) \cdot T^{-3}, \label{RelaxTime}
\end{equation}%
give the following relation for electron temperature $T_{ei}$
after passage of $i$ SFQ pulses:
\begin{equation}
T_{ei}^{2}=T_{ei-1}^{2}+2Q/\gamma _{e}\Lambda _{s}.
\label{TempJump}
\end{equation}
For the junction parameters in our example, the dissipated energy
is $Q\approx I_{c}\Phi _{0}\approx 2\cdot 10^{-20}$J, and the
shunt resistor volume is $\Lambda _{s}=4.5\cdot 10^{-18}$m$^{3}$.
If we start then from $T_{e0}=0$, electron temperature jumps
sequentially to 7 mK, 10 mK, 12 mK ... These figures show that our
shunt resistor will not be overheated during a "single shot"
experiments with only a few SFQ pulses.

As one can see from Eqs.~(\ref{HeatBal}) and (\ref{ElTemp}) with
$P_{J}=0$, and assuming again that $T_{p}=0$, electron temperature
changes between the jumps with time $t$ non-exponentially:
\begin{equation}
T_{e}^{3}=(\gamma _{e}/3\Sigma)/t\,  \label{Relaxation}
\end{equation}
and the notion of the relaxation time introduced in
Eq.~(\ref{RelaxTime}) can not be used rigorously, but gives only
the characteristic time scale of the temperature variations.
Nevertheless, Eq.~(\ref{RelaxTime}) shows qualitatively that
overheating of the shunt resistors becomes significant if the
pulse repetition rate exceeds 10 MHz. In particular, at an
achievable 10 GHz clock frequency the power is dissipated
quasi-continuously and electron temperature of the shunt is
constant and high, about 0.4 K.

From this discussion we see that the noise temperature $T_{N}$ of
a Josephson junction in our example can range from 75 mK for the
junction that is not switching frequently and has the bias
resistor of large volume, to 0.44 K for the continuously switching
junction with the small-volume bias resistor. However, even in the
regime of low $T_{N}$, the real electron temperature of the bias
resistor is high (about 0.4 K) and can produce other overheating
effects besides errors in the SFQ circuit operation.

\subsection{Phonon resistances and temperature distribution along
the chip}

The second important thermal resistance in the chain (Fig.~1) of
the heat propagation from a resistor, is that of the substrate. A
typical substrate can be viewed as a generic insulator crystal
with thermal conductivity $K$ that can be written at low
temperatures as
\begin{equation}
K=Cvl/3\, ,  \label{e2}
\end{equation}
and depends on three different parameters. Specific heat $C$ and
an average speed of sound $v$ are characteristics of the material
that are temperature-dependent but are practically independent of
material imperfections and sample geometry. For instance, for
silicon substrates at low temperatures these constants give:
\begin{equation}
K(l)=1200~ T^{3}l~ \lbrack W m^{-2} K^{-4}]\, . \label{e3}
\end{equation}
In contrast to $C$ and $v$, the mean free path $l$ depends
strongly on the crystal quality, doping concentration and other
parameters, and varies from several centimeters in single crystals
to few tens of nanometers in glasses. For instance, the mean free
path of thermal phonons in single-crystal Si can reach up to few
centimeters at sub-Kelvin temperatures \cite{Klitsner}. In this
case, the actual thermal conductivity and temperature profile in a
wafer with thickness $d\ll l$ is determined by properties of
phonon scattering at the surfaces; therefore thermal conductivity
depends on surface properties. In a typical situation of rough
surface with diffusive scattering, the conductivity can still be
estimated from Eqs.~(\ref{e2}) or (\ref{e3}) by taking $l\sim d$.
For specular reflection, ballistic phonon propagation in
single-crystal substrate can lead to complicated temperature
profiles determined by the ``geometric optics'' of phonons
\cite{Wichard}.

For a thin wafer, the temperature profile along it due to heat
spreading from a resistor is determined by the competition between
the heat conduction along the wafer and heat transfer to the
sample holder which acts as the heat sink. The heat resistance to
the sink consists (Fig.~1) of the heat resistance of the layer of
glue (epoxy) and the ``Kapitza'' resistance of the Si-epoxy and
epoxy-Cu interfaces due to mismatch of acoustic properties of
these materials. The acoustic-mismatch theory of the Kapitza
resistance \cite{little59,lounasmaa74} describes the interface
conductance $G_K$ in terms of probability $D$ for phonons to be
transmitted through the interface. In the case of plain interface
between the two materials with equal sound velocities $v$,
``transparency'' $D$ is determined by the difference of their
acoustic impedances $Z_{1,2}$:
\begin{equation}
D =4Z_1Z_2/(Z_1+Z_2)^{2} \, , \label{e4}
\end{equation}
where $Z_{i}=\rho _i v$, and $\rho_i$ is the mass density of the
material. The interface thermal conductance is then given by the
expression similar to Eq.~(\ref{e2}), $G_K=C v D/4$, i.e.
\begin{equation}
G_K =\gamma_{K} T^{3}\, . \label{e5}
\end{equation}
The values of coefficient $\gamma _{K}$ for various interfaces,
including those encountered in a typical SFQ chip (Fig.~1) can be
found, e.g., in \cite{lounasmaa74,KapRev,gmelin99}. In general,
$\gamma _{K}$ lies in the range $10 - 10^3$ Wm$^{-2}$K$^{-4}$ for
most of the dielectric-to-dielectric or dielectric-to-metal
interfaces.

Thermal conductance of the epoxy or other glue layer between the
substrate and the sink depends on several factors, including the
deposition method \cite{matsumoto77}. For sufficiently thick
layers, however, the conductance should follow the $T^2$
temperature dependence characteristic for amorphous materials in
the sub-kelvin temperature range \cite{lounasmaa74,pohl02}:
\begin{equation}
K=\gamma  T^{2}\, ,  \label{e6}
\end{equation}
where $\gamma$ is within the range $10^{-3} - 10^{-1}$
Wm$^{-1}$K$^{-3}$. The two thermal resistances (\ref{e5}) and
(\ref{e6}) are connected in series and both contribute to the
substrate-sink resistance. The dominant contribution is determined
by the thickness $d_{a}$ of the amorphous layer, with the
transition between mostly Kapitza to mostly bulk resistance
occurring at a characteristic value $d_{a}=\gamma/(\gamma _{K}T)$.
At $T\simeq$ 1~K, and $\gamma$ and $\gamma _{K}$ at intermediate
values within the ranges mentioned above, $d_{a}$ is order of 100
$\mu$m, implying that at the sub-kelvin temperatures the
substrate-sink resistance should typically be dominated by the
Kapitza resistance. This is because the thickness of the glue
layer, while uncertain, should not be much larger than 100 $\mu$m.

\begin{figure}[tb]
\begin{center}\leavevmode
\includegraphics[width=0.85\linewidth]{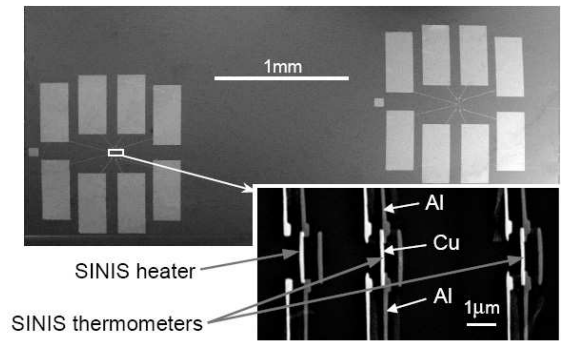}
\caption{Silicon wafer with SINIS heaters and thermometers used
for the overheating measurements. } \label{SampleTempProf}
\end{center}
\end{figure}

\begin{figure}[tb]
\begin{center}
\includegraphics[width=\linewidth]{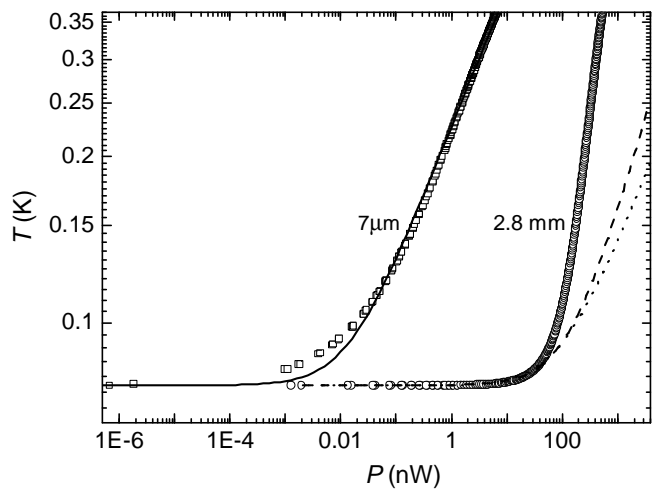}
\caption{Measured temperature on the surface of a silicon
substrate as a function of the heating power at 7 $\protect\mu$m
(squares) and 2.8 mm (circles) distances from the point-like
heater. The solid, dashed and dotted lines give the results of
theoretical modelling described in the text.} \label{TemProf}
\end{center}
\end{figure}

To obtain experimental insight in the actual temperature profile
on a standard silicon substrate we performed a few experiments
with a virtually point-like heater and local thermometers at
different distances from it. A small (0.5 $\mu$m x 0.5 $\mu$m)
superconductor-insulator-superconductor (SIS) tunnel junction
placed in the center of a silicon chip and biased above the double
gap voltage was used as a heat source. Superconductor electrodes
were aluminium, with aluminium oxide as the tunnel barrier.
Similar junctions at different distances from the heater were used
as thermometers. In some of the experiments
superconductor-insulator-normal metal-insulator-superconductor
(SINIS) structures were used instead as heaters and thermometers
with similar results (see Fig. \ref{SampleTempProf}). The size of
the normal copper island in this case was about 2 $\mu$m x 0.2
$\mu$m. In both schemes, the strong temperature dependence of the
quasiparticle current-voltage characteristics served as a local
probe of temperature on the surface of the chip. Figure
\ref{TemProf} shows results of a measurement on a wafer used by
HYPRES \cite{HYPRES} as a standard substrate for Josephson
junction circuits. The thickness of the boron doped (10
$\Omega$~cm), double-side polished $ \langle100\rangle$ wafer was
0.635 mm, and the size of the chip was 8 mm $ \times$ 8 mm.
Temperature as a function of power was measured at two different
distances (7 $\mu$m and 2.8 mm) from the heat source placed
approximately in the center of the chip. The bath temperature of
the experiment was 77 mK. The temperature at the distance of 7
$\mu$m from the heat source reaches twice the bath temperature at
the power level of 0.15 nW. Power of 180 nW is required to heat
the thermometer at the distance of 2.8 mm from 77 mK up to 150 mK.
In addition to the samples made from a typical HYPRES wafer, few
other silicon wafers with different thicknesses of oxide layer
(including wafer with thin native oxide layer) have been measured
and showed similar results.

To attempt fitting these results using the understanding of the
phonon heat transport described at the beginning of this section,
we need to use different models for the short and long distances
from the heater. For distances shorter than the substrate
thickness $d$, i.e. including the 7 $\mu$m in experiment, phonons
propagate ballistically from the point source in the Si substrate.
As a relatively crude but simple approximation, one can assume
that the point heater at the surface of the Si substrate radiates
the power $P$ uniformly in the hemisphere filled by the substrate.
In this case the energy density $u$ at a distance $r$ from the
source is
\begin{equation}
u= P/(2\pi r^2 v) \,  .\label{e8}
\end{equation}
A fraction $f$ of this energy in the non-equilibrium flux of
phonons is absorbed by the thermometer. In this process, it is
seen by the thermometer as the excess energy density $u_e$ that
corresponds to local equilibrium at some temperature $T$ above the
background bath temperature $T_0$. In the temperature range of the
experiment, we can use the usual Debye law, $u \propto T^4$, for
the equilibrium energy density of the phonon system,
\begin{equation}
u_e= (\nu/4) (T^4-T_0^4) \, , \label{e9}
\end{equation}
where $\nu$ is the coefficient in the Debye specific heat $C=\nu
T^3$. Equating the energy density (\ref{e9}) to a fraction $f$ of
density (\ref{e8}), we get the effective substrate temperature at
the distance $r$ from the source:
\begin{equation}
T(r, P)=[T_0^4+\frac{2f P}{\pi r^2 \nu v } ]^{1/4}\, . \label{e10}
\end{equation}

The 7-$\mu$m solid line in Fig.~\ref{TemProf} shows the dependence
of temperature $T$ on power $P$ calculated from Eq.~(\ref{e10})
using the fraction $f$ as a fitting parameter. The combination of
other factors $\nu v$ in Eq.~(\ref{e10}) is the same as the one
that determines the thermal conductivity in Eqs.~(\ref{e2}) and
(\ref{e3}), and its value for Si can be taken from Eq.~(\ref{e3}).
We see that one can obtain good fit of the observed $T(P)$
dependence with $f\simeq 0.72$.

The strength of the substrate heating at the larger distance, 2.8
mm, is determined by the interplay of the horizontal heat flow
along the substrate, and the heat leakage into the sink through
the glue layer (Fig.~1). As was discussed above, since the phonon
mean free path $l$ is much larger than the substrate thickness $d$
= 0.635 mm, the horizontal heat conductance $K_h$ is dominated by
the phonon scattering at the substrate surface. For mostly
diffusive scattering, $l\simeq d$, and Eqs.~(\ref{e2}) and
(\ref{e3}) give:
\[ K_h=d~K(d)  \equiv \sigma T^{3}\, . \]
The vertical heat conductivity $K_v$ is determined by either
Kapitza resistance or heat resistance of amorphous glue layer and
can be written as
\[ K_v=\lambda^{(\beta)} ~ (T^{\beta}-T_{0}^{\beta}),  \]
where the power $\beta$ is equal to 3 or 4, and the coefficients
$\lambda^{(\beta)}$ include all temperature-independent factors.
Neglecting the influence of the external boundaries of the
substrate we assume that the heat flow from the heater has radial
symmetry. In this case, equation describing the balance between
the horizontal and vertical heat flows has the form:
\begin{equation}
\frac{1}{r}\frac{\partial }{\partial r}(r \sigma
T^{3}\frac{\partial T}{\partial r})=\lambda^{(\beta)} ~
(T^{\beta}-T_{0}^{\beta}) \, , \label{e12}
\end{equation}
where $r$ is the radial distance from the heater. This equation is
valid on the scale of distances larger that the substrate
thickness $d$, and should be solved with the boundary conditions
describing the generation of power $P$ by the heater at $r=0$ and
negligible heat flow though the outer edge of the substrate at
$r=R$, ($R \simeq 4$ mm for the data presented in
Fig.~\ref{TemProf}):
\begin{equation}
\sigma T^{3}\frac{\partial T}{\partial r}+\frac{P}{2\pi r}=0\, ,
\;\; r\rightarrow 0\, ;  \;\;\; \frac{\partial T}{\partial r}=0\,
, \;\; r = R . \label{e13}
\end{equation}

Results of solution of Eq.~(\ref{e12}) with the boundary
conditions (\ref{e13}) are shown as dotted ($\beta$ = 4) and
dashed ($\beta$ = 3) lines for 2.8 mm in Fig.~\ref{TemProf}. In
these curves, an attempt was made to describe the data by fitting
$\lambda^{(\beta)}$. One can see that the model does not describe
the rapid temperature rise with the power $P$ in the whole range
of powers. The initial upturn of temperature with power can be
reproduced assuming either $\beta = 3$ or $\beta = 4$ if we take
\begin{equation}
\lambda^{(3)} =6\, \text{W}/\text{m}^2\text{K}^3\, , \;\;\;
\mbox{or}\;\;\; \lambda^{(4)} =44 \,
\text{W}/\text{m}^2\text{K}^4\, . \label{e14}
\end{equation}
The variation of modelling curves with $\beta$ (more rapid
temperature rise for $\beta =3$) suggests that the discrepancy
between theory and experiment is due to the substrate-sink heat
conductance dominated by the mechanism with weaker temperature
dependence, although it is unclear what could be such a mechanism
in our set-up. Nevertheless, the fact that the numbers in Eq.
(\ref{e14}) lie within the reasonable range discussed above, makes
it possible to say that the overall level of the substrate-sink
heat conduction agrees roughly with theoretical expectations.
Finally, the overall level of the substrate overheating presented
in Fig.~\ref{TemProf} seems to be consistent with other
measurements \cite{therm} on similar Si substrates, although the
limitations of the thermometer used in \cite{therm} do not alow
for detailed comparison.

\subsection{Thermalization of resistive films}

\begin{figure}[tb]
\begin{center}\leavevmode
\includegraphics[width=0.75\linewidth]{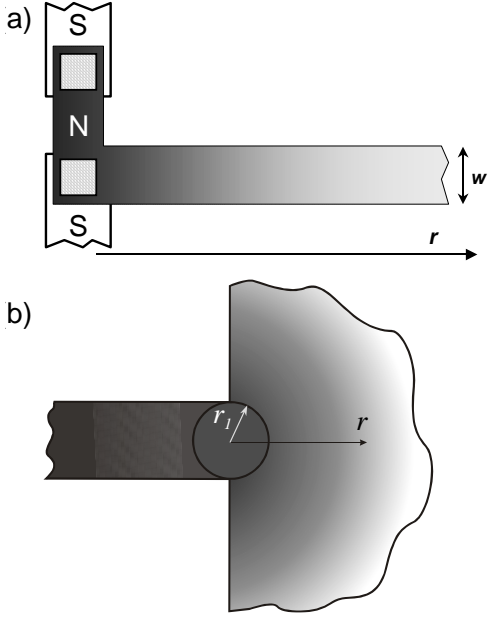}
\caption{The sketch of the shunt resistor with cooling fins of
different geometry.} \label{CoolFin1}
\end{center}
\end{figure}

In some cases gradient of electron temperature along the resistor
could play a significant role in electron-phonon relaxation
discussed above. In particular, this is the case if the resistor
consists of two parts. One (the vertical strip in
Fig.~\ref{CoolFin1}(a)) actually serves as the resistor, while the
other (the horizontal strip) does not carry electric current and
serves as the cooling sink or fin. This shape of the resistor
enables one to optimize the two parts separately simplifying the
design procedure.
In this subsection, we estimate the length which limits the useful
size of the cooling fin. Increasing the fin size beyond this
length does not improve electron cooling because of the finite
electron thermal conductivity in the resistor film. The
distribution of electron temperature along the fin is determined
by the electron thermal conductivity $K_{e}$ in the film and
strength of electron-phonon relaxation $K_{e-p}$. The conductivity
$K_{e}$ is proportional to the film thickness $d_f$ and electron
temperature $T_{e}$:
\begin{equation}
K_{e}=d_{f} \kappa _{e} T_{e}=\sigma _{e} T_{e}, \label{radial}
\end{equation}
whereas
\begin{equation}
K_{e-p}=d_{f} \Sigma (T_{e}^{5}-T_{0}^{5})=\lambda
_{e-p}(T_{e}^{5}-T_{p}^{5}).
\end{equation}

A similar model of the temperature distribution is obtained by
assuming a uniform semi-infinite film connected at its side to a
hot spot (see Fig.~\ref{CoolFin1}(b)). This hot spot approximates
one end of a resistor of width $2r_{1}$. Equations corresponding
to the distribution of electron temperature along the fin both in
the linear ($m$ = 0) and in cylindrical ($m$ = 1) case can be
presented as:
\begin{equation}
\frac{1}{r^{m}}\frac{\partial }{\partial r}r^{m}\sigma _{e}T_{e}\frac{%
\partial T_{e}}{\partial r}=\lambda _{e-p}(T_{e}^{5}-T_{p}^{5}).
\label{radial4}
\end{equation}
Equation (\ref{radial4}) for the linear case ($m = 0$) and
$T_{e}>>T_{p}$ has an analytical solution:
\begin{equation}
T_{e}=(\xi_{S}/(r+r_0))^{2/3} ,  \label{Tevsr}
\end{equation}
where $\xi_{S}\equiv \sqrt{(14/9)~\kappa_{e}/\Sigma }$, and $r_0=
\xi_S/T_{e1}^{3/2}$ with $T_{e1}$ denoting electron temperature at
the left (hot) boundary of the fin. Solution for the cylindrical
case ($m = 1$) is qualitatively similar if $r$ is replaced with
$r-r_1$.

The distance $r_{d}$ at which the efficiency of electron-phonon
relaxation $\thicksim T_{e}^{5}$ becomes two times smaller than at
the boundary ($T_{e}^{5}$ = 0.5 $T_{e1}^{5})$,
\begin{equation}
r_{d}=(2^{3/10}-1) r_0 \cong 0.23 r_0 ,
\end{equation}%
can be considered as the maximum effective size of the fin:
increase of the fin size beyond $r_{d}$ does not noticeably
improve resistor cooling. On the other hand, for $r \leq r_{d}$,
one can neglect the variation of electron temperature in the fin,
and its thermal resistance is determined by its volume $\Lambda$
and electron phonon coupling constant $\Sigma$ through
Eq.~(\ref{ElTemp}). As a numerical example, we take a copper film
for which $\kappa _{e}\simeq 1$ WK$^{-2}$m$^{-1}$ and $\Sigma
\simeq 2\cdot 10^{9}$ WK$^{-5}$m$^{-3}$. This gives $r_{d}\sim $ 2
mm at the electron temperature of $T_{e}=$ 100 mK.

\section{General recommendations about thermal design}

\begin{figure}[t]
\begin{center}\leavevmode
\includegraphics[width=0.75\linewidth]{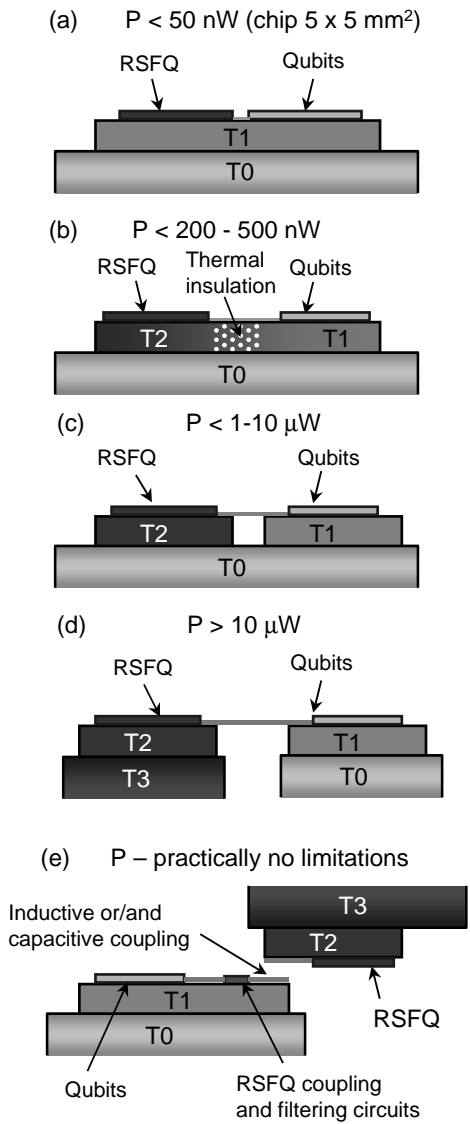}
\caption{Optimization of thermal design for the circuits of
different complexity. Designs with both circuits mounted at the
same holder: (a) RSFQ and quantum circuits on the same Si chip,
(b) substrate with improved thermal insulation between two parts,
(c) two-chips solution. Separate cooling of circuits with
different temperatures and power dissipations: two chips with
independent cooling connected by RF lines (d) and inductively (or
capacitively) coupled (e).
 }\label{ThDes}\end{center}\end{figure}

To summarize our arguments we present in Fig.~\ref{ThDes} the
thermal designs for the SFQ-qubit circuits with different levels
of power dissipation. A scientific experiment on qubits controlled
by a simple SFQ circuit with the power dissipation below 50 nW is
basically doable on a single silicon chip (Fig.~\ref{ThDes}(a)).
In this case, the temperature of silicon substrate
($T_{\mathrm{1}}$) in the vicinity of qubits can only slightly
exceed the bath temperature $T_{\mathrm{0}}$ while the electron
temperature of resistors of the SFQ circuit can be as high as 500
mK. If needed, electron temperature in a few resistors can be
reduced to about 100 mK by cooling fins. A higher dissipation
power is acceptable for the chip with additional thermal
insulation (for example, porous silicon or specially etched
structure on the back of the chip) between areas with qubits and
SFQ circuits (Fig.~\ref{ThDes}(b)). The relatively high thermal
resistance along the substrate as compared to the resistance
between the substrate and the sample holder makes it possible to
keep qubits at low temperature ($T_{\mathrm{1}}$). The increase of
power dissipation above 500 nW requires even better thermal
separation of the circuits, and in this case the SFQ circuits
could not be placed on the same chip with the qubits. The two-chip
design (Fig.~\ref{ThDes}(c)-(e)) practically eliminates the
problem of overheating of qubit circuitry. Moreover, it allows
utilization of two independent fabrication technologies for SFQ
and qubit circuits, and as a result, the conventional SFQ circuits
can be immediately used in qubit support circuits. Both chips can
be kept on the same metal sample holder (Fig.~\ref{ThDes}(c)) if
the dissipated total power does not increase $T_{0}$ significantly
above the temperature of the mixing chamber. Usually the cooling
power of a dilution refrigerator is not a problem up to a
dissipation level of few $\mu$W. For higher power (more
complicated SFQ circuits) the separate active cooling of both
circuits is required (Fig.~\ref{ThDes}(d)). The quantum circuit is
supposed to be at the temperature below 50 mK, but the SFQ chip
can be kept at higher temperature ($T_{\mathrm{3}}$). In this case
few different refrigerating stages with the cooling power in the
mW range can be used for cooling SFQ circuits. As discussed above,
the electron temperature of shunt resistors in the SFQ circuits
with reduced critical current should typically be about 500 mK and
does not strongly depend on the lattice temperature which is below
500 mK. A $^{3}$He evaporation refrigerator delivering enough
cooling power at a temperature of about 300 mK is a very
attractive solution for cooling SFQ circuits. Unfortunately this
leads to essential complication of cryogenic equipment. A more
natural solution is to make use of different stages in the
dilution refrigerator for cooling the SFQ circuit: 1 K pot
(temperature 1-2 K) or $^{3}$He evaporator (600 - 800 mK). The
latter is more attractive and natural due to lower temperature and
because some heating of this stage is in any case necessary for
operation of the dilution refrigerator.

When we consider two circuits mounted at different temperatures,
the thermal load through connecting wires should be limited to
prevent overheating of the qubit circuit. In the case of
superconducting (Nb, Al) leads connecting two chips at 1 K and at
20 mK, respectively, a heat load is about 100 nW, if the total
cross-section area of the wires is 0.1 mm$^{2}$ and the length is
10 mm. This may be acceptable in most cases, but an inductive
or/and capacitive coupling between the circuits (Fig.~\ref{ThDes}
(e)) is a more suitable option for a fully scalable solution.

Our discussion assumed a chip size 5 $\times$ 5 mm$^{2}$ glued to
a bulk copper heat sink, and the power levels mentioned above
should be considered as order-of-magnitude estimates. These
estimates can be affected by changes in the system design, i.e.,
different substrate materials and geometry (both area and
thickness), improved thermal contact between the substrate and the
heat sink (increased contact area, modification of the contact
surfaces, etc.), and more powerful dilution refrigerator.

\section{SFQ circuit operating at sub kelvin temperatures}

The direct way of transferring SFQ circuit design to the
sub-kelvin temperature range is to reduce all currents, including
critical currents of Josephson junctions and dc bias currents,
proportionally to the effective temperature of the junctions. With
this scaling, the existing technology can then be used as such and
most of the existing SFQ logic elements can be adopted with some
modifications. We carried out such scaling for several basic SFQ
circuits, and present here the results for the simplest circuit, a
balanced comparator. The balanced comparator is one of the
principal building blocks of more complex circuits, but it can
also be used directly as a thermometer, providing a convenient way
of testing thermal characteristics of the circuit.

\begin{figure}[t,b]
\begin{center}\leavevmode
\includegraphics[width=0.8\linewidth]{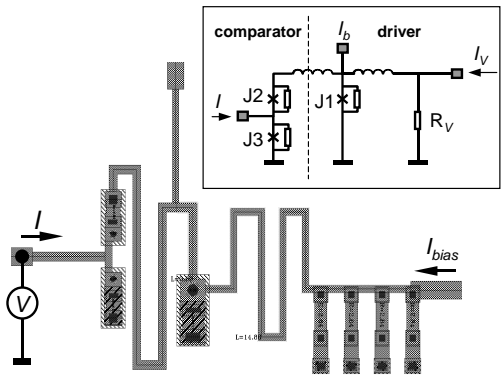}
\caption{Layout and equivalent circuit of the measured comparator.
} \label{Comp}
\end{center}
\end{figure}

\begin{figure}[t,b]
\begin{center}\leavevmode
\includegraphics[width=0.80\linewidth]{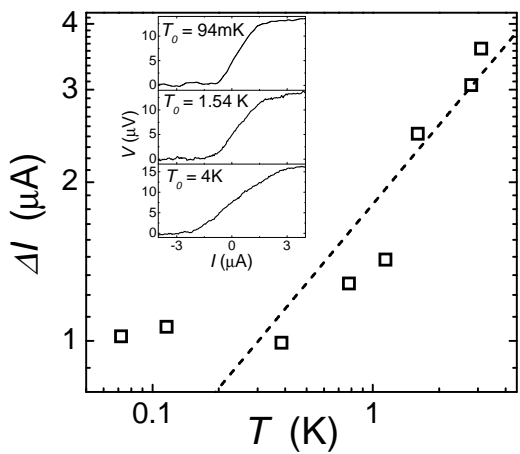}
\caption{Temperature dependence of the width of the gray zone of
the measured comparator. Squares and dashed line are experimental
data and theoretical prediction for thermal limit, respectively.}
\label{GrZon}
\end{center}
\end{figure}

The circuit (Fig.~\ref{Comp}) contains three shunted Josephson
junctions and  requires for its operation three dc bias currents.
One bias current provides the necessary dc bias of comparator
junctions J2 and J3. The junctions are in superconducting state
and carry dc current about 70\% of their critical currents. The
other bias current is applied to a relatively low resistance $R_V$
to supply a low, e.g. 12$\mu $V, voltage drop $V_{in}$. Junction
J1 converts this dc voltage into a sequence of SFQ pulses
generated with frequency $f_{in} = (2e/h) V_{in}$. The SFQ pulses
escape from the circuit either via junction J2 or via junction J3
depending on the dc current $I$. Thermal or quantum noise smoothes
the otherwise sharp transition between these two escape channels.
According to a well-developed and experimentally confirmed theory,
the current width $\Delta I$ of the ``gray zone'' of this
transition in the regime dominated by thermal fluctuations (see,
e.g., \cite{rsfq, semenov:3617}) is:
\begin{equation}
\Delta I \simeq(2\pi\alpha I_{T}I_{C})^{1/2}. \label{GrZ}
\end{equation}
Here $I_{T}=2\pi k_{B}T/\Phi_{0}$ and $\alpha$ is a dimensionless
parameter determined by comparator and driver characteristics -
see \cite{Filippov:2240, Filippov:2452, semenov:3617} for details.
Temperature dependence of $\Delta I$ can be used to determine
electronic temperature in the shunt resistors.

The comparator was fabricated using standard 100 A/cm$^{2}$ Nb
trilayer process of HYPRES with PdAu resistors. The layout of the
comparator, which includes two nominally identical junctions (left
part of the circuit) and the driver, is shown in Fig.~\ref{Comp}.
The comparator parameters, $I_{c}$ = 10 $\mu$A, $R_{s}$ = 2
$\Omega$, were chosen for operation at sub-kelvin temperatures.
The junction critical current $I_{c}$ was thus reduced by an order
of magnitude from its usual value for temperatures around 4 K. For
these parameters, the junctions are overdamped, and the crossover
temperature $T^*$ between the regimes of thermal and quantum
fluctuations \cite{weiss} is given by the relation $T^* \simeq
eV_c/\pi k_{B} \simeq$  70 mK, where $V_c=I_{c}R_{s}$. This means
that quantum broadening of the gray zone of the comparator can be
neglected in our measurements.

The measurement procedure was similar to that described in
Ref.~\cite{semenov:3617}: the width $\Delta I$ of the gray zone
was obtained from the dc voltage $V$ across one of the comparator
junctions as a function of the applied current $I$ (inset in
Fig.~\ref{GrZon}) and its temperature dependence is presented in
Fig.~\ref{GrZon}. The dashed line corresponds to the theoretical
prediction in the thermal limit [Eq.~(\ref{GrZ})] assuming that
the effective electron temperature of the resistors coincides with
the bath temperature $T$. At bath temperatures above 0.4 K
experimental behavior of $\Delta I$ agrees well with the
theoretical prediction. At lower temperatures, $\Delta I$ does not
demonstrate noticeable dependence on $T$ as a result of the
overheating of the comparator circuit. This prevents the reduction
of electron temperature below 0.4 K.

Electrical and geometrical parameters of the circuit can be used
to estimate the electron temperature expected from the heating
model discussed in Sec.~II. The circuit contains four resistors.
Two shunt resistors of the comparator junctions are located about
4 $\mu$m from the junctions. Each of them occupies an area 7.25
$\mu$m $\cdot $12.5 $\mu$m $\simeq$ 90 $\mu $m$^{2}$. One more
resistor with the area 10.5 $\mu $m $\cdot $ 14.5 $\mu $m
$\sim$152 $\mu $m$^{2}$ shunts the driver junction J1 and is
located at about 50 $\mu $m from the comparator junctions. The
last resistor $R_V$ (seen in the layout of Fig.~\ref{Comp} as 4
resistors in parallel) has the total area 25.5 $\mu$m $\cdot $ 6
$\mu$m x 4 $\simeq$ 612 $\mu $m$^{2}$ and is 125 $\mu $m away from
the comparator. All resistors are made of 0.1 $\mu $m PtAu film
with 2 $\Omega$ sheet resistance, and designed values of the
comparator shunts, driver shunt, and resistor $R_{V}$ are 2
$\Omega$, 1.8 $\Omega$, and 1.4 $\Omega$, respectively. The
voltage drop across the generator shunt, $R_{V}$ resistor, and the
cumulative voltage drop on both comparator shunts are all about 12
$\mu $V, while the distribution of the voltage between the two
comparator junctions J2 and J3 depends on the current $I$.

We estimate electron temperature $T_{e}$ in the shunts of the
comparator junctions at the center of the gray zone ($I\simeq 0$),
when both shunts have the same voltage drop 6 $\mu $V. The shunt
parameters from the previous paragraph give their volume $\Lambda
=9\cdot 10^{-18}$ m$^{3}$ and the dissipated power $P=1.8\cdot
10^{-11}$ W. Using these numbers together with negligible phonon
temperature $T_{p}$ and an estimate of electron-phonon constant
(\ref{Sigma}) in Eq.~\ref{ElTemp}, we get $T_{e}\simeq 0.3$~ K.
Remote resistors do not affect the estimate of $T_e$. Their
electrical noise changes only the comparator bias and does not
contribute to the width of the gray zone in the case of identical
junctions J2 and J3. The phonon overheating in the vicinity of the
comparator junctions due to the power dissipated in these
resistors can be estimated from Eq.~(\ref{e10}) to be negligible,
$\sim 30$ mK. This means that electron overheating in the junction
shunt resistors is indeed the dominant overheating factor in our
experiment. We note that the measured overheating is somewhat
higher than the estimate within our model. The most probable
reason for this discrepancy is a small asymmetry between the
junctions J2 and J3 which makes it possible for the noise of all
resistors in the circuit to contribute to $\Delta I$. In view of
this and other possible sources of extra broadening, agreement
between the estimate of $T_e$ and the observed saturation of
$\Delta I$ is quite good.

\section{Conclusion}

We have analyzed thermal properties of a typical SFQ circuit at
sub-kelvin temperatures and performed several measurements testing
the basic elements of the heat conduction scheme (Fig.~1) of the
circuit. Local overheating of electrons in resistors is controlled
by electron-phonon coupling, while global overheating of the chip
is determined by the competition between ballistic phonon
propagation along the substrate and the leakage into the heat sink
that is limited by the Kapitza resistance or the thermal
resistance of the glue layer. Our analysis and data suggest that
integration of simple SFQ circuits with qubits on a single chip
should be possible if the total power dissipated by the SFQ
components is below 50 nW. Scalable solutions for a multi-qubit
system with large power dissipation require two-chip (hybrid)
designs with separate active cooling of the qubit and the SFQ
chips. An alternative strategy based on complete revision of the
SFQ approach (e.g., development of reversible SFQ circuits) should
be considered as a longer term goal.

\section{Acknowledgment}

We thank Yu. Polyakov for his help with grey zone measurements.
This work was supported in part by "RSFQubit" FP6 project of
European Union (A.M.S and J.P.P.), and in part by ARDA and DOD
under the DURINT grant \# F49620-01-1-0439 and by the NSF under
grant \# EIA-0121428 (D.V.A. and V.K.S.). Integrated circuit with
Josephson comparator has been fabricated at HYPRES, Inc.

\end{document}